%% file: paper_jrn_vmf_scf.tex
\documentclass[journal,comsoc]{IEEEtran}

\usepackage[T1]{fontenc}

\usepackage[pdftex]{graphicx}
\graphicspath{{./figs/}{./figs/bio/}}

\usepackage[acronym]{glossaries}
\loadglsentries[acronym]{acronyms}

\usepackage{amsmath}
\providecommand{\abs}[1]{\lvert#1\rvert} 
\providecommand{\norm}[1]{\left\lVert#1\right\rVert} 

\providecommand{\mean}[1]{\mathbb{E}\left\{#1\right\}} 

\DeclareMathOperator{\sinc}{sinc} 

\usepackage{color}

\ifCLASSOPTIONcompsoc
 \usepackage[caption=false,font=normalsize,labelfont=sf,textfont=sf]{subfig}
\else
 \usepackage[caption=false,font=footnotesize]{subfig}
\fi

\usepackage{multirow}

\newcommand\copyrighttext{%
  \textcopyright \the\year{} IEEE.
  Personal use of this material is permitted. \\
  Permission from IEEE must be obtained for all other uses, including reprinting/republishing this material for advertising or promotional purposes, collecting new collected works for resale or redistribution to servers or lists, or reuse of any copyrighted component of this work in other works.
}

\begin{document}

\title{Correlation Properties in Channels with \\ von Mises-Fisher Distribution of Scatterers} 

\author{Kenan~Turbic,~\IEEEmembership{Member,~IEEE,}
  Martin~Kasparick,
  and S\l{}awomir~Sta\'{n}czak,~\IEEEmembership{Senior Member,~IEEE}%
\thanks{%
%
This work was partially supported by the Federal Ministry of Education and Research (BMBF, Germany) in the “Souverän. Digital. Vernetzt.” programme, joint Project 6G-RIC; project identification numbers: 16KISK020K and 16KISK030.
It was developed within the scope of COST Action CA20120 (INTERACT).
The associate editor coordinating the review of this paper and approving it for publication was XX. 
(\textit{Corresponding author: Kenan Turbic.})
}%
\thanks{The authors are with the Wireless Communications and Networks Department, Fraunhofer Institute for Telecommunications, Heinrich Hertz Institute (HHI), Berlin, 10587 Germany (e-mail: kenan.turbic@hhi.fraunhofer.de, martin.kasparick@hhi.fraunhofer.de, slawomir.stanczak@hhi.fraunhofer.de).
}
\thanks{S. Stanczak is also with Technische Universit{\"a}t Berlin, Berlin, Germany.}%
}





\IEEEpubid{%
  \begin{minipage}{0.8\textwidth}\ \\[10mm] \centering
      \footnotesize \copyrighttext
  \end{minipage}
}

\maketitle

\begin{abstract}
This letter presents simple analytical expressions for the spatial and temporal correlation functions in channels with \acrlong{vmf} scattering.
In contrast to previous results, the expressions presented here are exact and based only on elementary functions, clearly revealing the impact of the underlying scattering parameters.
The derived results are validated by a comparison against numerical integration, where an exact match is observed.
To demonstrate their utility, the presented results are used to analyze spatial correlation across different antenna array geometries and to investigate the temporal correlation of a fluctuating radar signal from a moving target.
\end{abstract}


\glsresetall

%
\IEEEpeerreviewmaketitle

\section{Introduction}
\label{sec:intro}
\IEEEPARstart{T}{he} \gls{vmf} distribution is a widely adopted statistical model for angular distribution of scatterers in wireless communication channels, e.g., \cite{Wang2023, Pizzo2022a}, taking both the azimuth and elevation aspects of propagating waves into account.
The good fit to measurements and the ability to model non-isotropic scattering with different degrees of angular concentration of multipath components are its main advantages over the alternative models in the literature \cite{Mammasis2009}.
Due to its simplicity, the \gls{vmf} distribution is preferred over more general models, such as the \gls{fb5} distribution \cite{Kent1982}. \looseness=-1

While the \gls{vmf} model is widely adopted, the associated statistics and the impact of the underlying parameters are not fully understood yet, primarily due to the lack of closed-form analytical results.
The existing studies of spatial correlation characteristics are either based on numerical integration \cite{Mammasis2009}, employ spherical wave expansion involving expressions with infinite sums of spherical Bessel functions \cite{Mammasis2010}, or obtain approximate analytical solutions under restrictive assumptions \cite{Zhu2019}.
While these studies have greatly contributed to the initial understanding of non-isotropic scattering channels, the provided expressions are cumbersome and prohibit analytical inference, or their applicability is very restricted.

In this letter we present simple closed-form expressions for spatial- and auto-correlation functions in \gls{vmf} scattering channels.
In contrast to previous results, the presented expressions involve only elementary functions and do not rely on any approximations.
Their simplicity facilitates analysis and clearly reveals the impact of the \gls{vmf} distribution parameters.

Following our earlier internal report on this work in \cite{TurbicCOST2024}, some similar derivation steps appear to be independently applied in \cite{Zeng2024}, considering channels between unmanned aerial vehicles employing uniform linear antenna arrays.
In contrast to this work, we present simple expressions applicable to arbitrary 3D antenna array geometries, identify the direct impact of \gls{vmf} scattering parameters, recognize the classic result for isotropic scattering as a special case and demonstrate the utility of the derived expressions in different use cases.
As a novel application of the \gls{vmf} scattering model, we additionally employ it to represent radar targets and analyze the impact of the target size on the corresponding signal fluctuation dynamics. \looseness=-1

The rest of this letter is structured as follows.
Channel model is introduced in Section~\ref{sec:ch_mod}, while Section~\ref{sec:corr} presents the derivation of the spatial and temporal correlation functions.
In Section~\ref{sec:results}, the obtained results are validated and used to analyze spatial correlation across different antenna arrays and to investigate temporal correlation of a signal from a moving radar target.
The paper is concluded in Section~\ref{sec:conclusions}.

\section{Channel Model}
\label{sec:ch_mod}
With multipath propagation taking place between a static \gls{tx} and a mobile \gls{rx}, the wideband channel transfer function can be written as
\begin{align}
  H(\Delta \mathbf{r}, f)
  =\sum_{n=1}^{N} A_n e^{-j2\pi f \tau_0^n} e^{j\frac{2\pi}{\lambda} \hat{\mathbf{k}}_n^T \Delta\mathbf{r}}
  \label{eq:ch}
\end{align}
where
\begin{description}
	\item[$f$] frequency;
	\item[$\lambda$] corresponding wavelength;
	\item[$N$] number of multipath components;
	\item[$A_n$] their amplitudes;
	\item[$\tau_0^n$] initial propagation delays;
  \item[$\hat{\mathbf{k}}_n$] \gls{doa} unit vectors, i.e.
  \begin{align}
    \hat{\mathbf{k}}_n = (\cos\phi_n\cos\psi_n, \;\sin\phi_n\cos\psi_n, \;\sin\psi_n)^T
  \end{align}
  \item[$\phi_n$] \glspl{aaoa};
  \item[$\psi_n$] \glspl{eaoa};
	\item[$\Delta \mathbf{r}$] antenna position relative to the reference point, e.g., position at the beginning of the channel observation;
  \item[$(\,.\,)^T$] vector transpose operation.
\end{description}
%

\IEEEpubidadjcol

For the convenience of the derivations in the following section, the channel evolution is considered as a function of the spatial displacement of the \gls{rx} relative to its arbitrary initial position.
The more traditional representation, as a function of time, follows from the adopted \gls{rx} mobility model, providing the time-displacement mapping.

The multipath components are assumed to arrive at the \gls{rx} with similar amplitudes, none of them being dominant.
According to the central limit theorem, the envelope of \eqref{eq:ch} exhibits Rayleigh distribution in this case.
While this requires $N\rightarrow\infty$ in the theoretical reference model, it was demonstrated that $N \geq 10$ is practically sufficient \cite{Patzold2012book}.

Scattering is assumed to occur in the far field of the mobile antenna, such that planar wave propagation can be assumed.
This further implies that the \glspl{doa} and amplitudes of the arriving multipaths can be assumed constant over the local areas, several tens or hundreds of wavelengths in size \cite{Molisch2011book}. \looseness=-1

The initial propagation delays, due to apparently random propagation path lengths at the reference point ($\Delta \mathbf{r} = 0$), result in random phase shifts of the components arriving at the \gls{rx}.
In practice, the differences between the delays $\tau_0^n$ are typically much larger than the period of the carrier \cite{Patzold2012book}.
Thereby, the corresponding initial phases observed at any given frequency can be modeled as uniform random variables, i.e., $\varphi_0^n \sim \mathcal{U}(0, 2\pi)$, which is the standard approach for narrowband channels%
\footnote{The transmission coefficient of a narrowband channel is obtained by evaluating the channel transfer function \eqref{eq:ch} at the considered carrier frequency.}.

The \glspl{doa} are modeled by the \gls{vmf} distribution, with its \gls{pdf} given by \cite{Mardia2000book}
\begin{align}
	p_{\phi\psi}(\phi, \psi)
  &=
  \frac{\kappa \cos\psi}{4\pi\sinh{\kappa}}
  e^{\kappa\left[ \cos\mu_\psi \cos\psi \cos(\phi - \mu_\phi) + \sin\mu_\psi \sin\psi\right]}
	\label{eq:vmf_pdf}
\end{align}
where
\begin{description}
	\item[$\mu_\phi$] mean \gls{aaoa};
	\item[$\mu_\psi$] mean \gls{eaoa};
	\item[$\kappa$] spread parameter.
\end{description}
The parameters $\mu_\phi$ and $\mu_\psi$ specify the main direction of concentrated scattering and $\kappa$ specifies the degree of concentration.
For $\kappa=0$, the distribution becomes uniform over a unit sphere.

\section{Spatial and Temporal Correlation}
\label{sec:corr}

\subsection{Spatial Correlation Function}
\label{sec:corr_scf}
\gls{scf} of the complex baseband envelope, normalized to unit power, is obtained as%
\footnote{We should note that the \gls{scf} is frequency-dependent, as imposed by the presence of the wavelength $\lambda$ in \eqref{eq:scf_gen0}.}
\cite{Patzold2012book}
\begin{align}
  \hat{R}_{hh}(\mathbf{d}_{\Delta})
  &= \mean{H^{*}(\Delta \mathbf{r}_1, f), H(\Delta \mathbf{r}_2, f)} \\
  &= \mean{e^{j\frac{2\pi}{\lambda} \,\hat{\mathbf{k}}_n^T \mathbf{d}_{\Delta}}}
  \label{eq:scf_gen0}
\end{align}
where
\begin{description}
	\item[$\mathbf{d}_{\Delta}$] spatial displacement vector, i.e. $\mathbf{d}_{\Delta} = \Delta \mathbf{r}_2 - \Delta \mathbf{r}_1$.
\end{description}
By using the \gls{pdf} in \eqref{eq:vmf_pdf}, the expectation \eqref{eq:scf_gen0} becomes
\begin{align}
	&\hat{R}_{hh}(\mathbf{d}_{\Delta}) = \frac{\kappa}{4\pi \sinh\kappa} \times \notag\\
	&\hspace*{2mm} \int_{-\pi/2}^{\pi/2} \left\{ \int_{-\pi}^{\pi} e^{\left( b_x\cos\phi+b_y\sin\phi \right) \cos\psi} d\phi \right\} e^{b_z\sin\psi} \cos\psi d\psi
	\label{eq:scf_gen}
\end{align}
with
\begin{align}
	b_x &= \kappa \cos{\mu_\phi}\cos{\mu_\psi} + j\frac{2\pi}{\lambda} d_{\Delta}^{x}
	\label{eq:b_x}
  \\
	b_y &= \kappa \sin{\mu_\phi}\cos{\mu_\psi} + j\frac{2\pi}{\lambda} d_{\Delta}^{y}
	\label{eq:b_y}
  \\
	b_z &= \kappa \sin{\mu_\psi} + j\frac{2\pi}{\lambda} d_{\Delta}^{z}
	\label{eq:b_z}
\end{align}
and $d_{\Delta}^{x/y/z}$ are components of the displacement vector $\mathbf{d}_{\Delta}$.

Employing \cite[Eq.~3.338.4]{Gradshteyn2007book} to solve the inner integral in the azimuth angle variable yields
\begin{align}
	&\hat{R}_{hh}(\mathbf{d}_{\Delta}) \notag\\
  &= \frac{1}{2} \frac{\kappa}{\sinh\kappa} \int_{-\pi/2}^{\pi/2} I_0\left( \sqrt{b_x^2+b_y^2} \cos\psi \right) e^{b_z\sin\psi} \cos\psi d\psi
\end{align}
by considering that $\cos{\psi}\geq0$ for $\abs{\psi} \leq \pi/2$.
From the relationship between the modified and the regular Bessel functions \cite[Eq.~8.406.3]{Gradshteyn2007book}, it follows
\begin{align}
	&\hat{R}_{hh}(\mathbf{d}_{\Delta}) \notag\\
	&\hspace*{2mm}= \frac{1}{2} \frac{\kappa}{\sinh\kappa} \int_{-\pi/2}^{\pi/2} J_0\left( \sqrt{B} \cos\psi \right) e^{b_z\sin\psi} \cos\psi d\psi
\end{align}
with
\begin{align}
  B
  &= \left(\frac{2\pi}{\lambda}\right)^2 \norm{\mathbf{d}_{\Delta}}^2 - \kappa^2\cos^2{\mu_\psi} - j 2 \kappa\cos{\mu_\psi} \, \hat{\mathbf{k}}_{\mu}^T \mathbf{d}_{\Delta}
  \label{eq:B}
\end{align}

where
\begin{description}
	\item[$\mathbf{k}_\mu$] mean \gls{doa} unit vector, i.e.
	\begin{align}
    \mathbf{k}_\mu = (\cos\mu_\phi\cos\mu_\psi, \sin\mu_\phi\cos\mu_\psi, \sin\mu_\psi)^T
    \label{eq:doa_mean}
  \end{align}
\end{description}

With a simple substitution, this integral can be reduced to the form in \cite[Eq.~6.616.5]{Gradshteyn2007book}, to obtain
\begin{align}
	\hat{R}_{hh}(\mathbf{d}_{\Delta})
	&= \frac{\kappa}{\sinh\kappa} \frac{\sinh\sqrt{b_z^2 - B}}{\sqrt{b_z^2 - B}}
	\label{eq:scf_sinh}
\end{align}
Finally, replacing \eqref{eq:b_z} and \eqref{eq:B} in \eqref{eq:scf_sinh}, and employing \cite[Eq.~1.311.2]{Gradshteyn2007book}, after some manipulation, yields
\begin{align}
  \hat{R}_{hh}(\mathbf{d}_{\Delta})
  &=
  \frac{\kappa}{\sinh\kappa}
  \mathrm{sinc}\left( \sqrt{\left(\frac{2\pi}{\lambda}\right)^2 \norm{\mathbf{d}_{\Delta}}^2 - \kappa^2 - j \frac{4\pi\kappa}{\lambda} \, \hat{\mathbf{k}}_{\mu}^T \mathbf{d}_{\Delta} } \right)
  \label{eq:scf_vmf}
\end{align}

As we observe, the \gls{scf} depends on both the displacement distance and the displacement direction relative to the mean \gls{doa}, with the dependence on the latter being controlled by the spread parameter $\kappa$.
For $\kappa=0$, the dependence on the displacement direction disappears%
\footnote{In the limit as $\kappa\rightarrow0$, the normalization factor in front of the $\mathrm{sinc}$ function in \eqref{eq:scf_vmf} is equal to one, as can be shown by applying L'Hopital's rule. Formally, a continuous extension of the function \eqref{eq:scf_vmf} is considered for $\kappa=0$.}
and the \gls{scf} reduces to the classical result for 3D isotropic scattering \cite{Cook1955}, i.e.
\begin{align}
  \hat{R}_{hh}(\mathbf{d}_{\Delta})
  &=
  \mathrm{sinc}\left( \frac{2\pi}{\lambda} \norm{\mathbf{d}_{\Delta}} \right)
  \label{eq:scf_iso}
\end{align}
The \gls{scf} is real in this case.
However, the \gls{scf} in \eqref{eq:scf_vmf} is complex in general, with its real part corresponding to the auto-correlation of the quadrature components and the imaginary one to their cross-correlation \cite{Patzold2012book}.

\subsection{Auto-correlation Function}
\label{sec:corr_acf}
By considering the spatial displacement between positions of the mobile antenna along its motion path, with employment of an appropriate mobility model, the \gls{scf} in \eqref{eq:scf_vmf} yields \gls{acf}.
For a linear motion at a constant velocity, the displacement vector can be expressed as
\begin{align}
  \mathbf{d}_{\Delta} &= \mathbf{v} \, \Delta t
  \label{eq:disp_vec_motion}
\end{align}
where
\begin{description}
  \item[$\Delta t$] time offset between \gls{rx} signal samples;
  \item[$\mathbf{v}$] mobile antenna velocity vector, i.e.
  \begin{align}
    \mathbf{v} = v (\cos\phi_v\cos\psi_v, \;\sin\phi_v\cos\psi_v, \;\sin\psi_v)^T
  \end{align}
  \item[$v$] mobile antenna speed;
  \item[$\phi_v$] motion direction azimuth;
  \item[$\psi_v$] motion direction elevation.
\end{description}
Replacing \eqref{eq:disp_vec_motion} in \eqref{eq:scf_vmf} yields the \gls{acf}, i.e.
\begin{align}
  \hat{R}_{hh}(\Delta t)
  &= \frac{\kappa}{\sinh\kappa}
  \mathrm{sinc}\left( \sqrt{\left(2\pi f_m \Delta t \right)^2 - \kappa^2 - j 4\pi\kappa f_\mu \, \Delta t } \right)
  \label{eq:acf_vmf}
\end{align}
where
\begin{description}
  \item[$f_m$] maximum Doppler frequency shift, i.e., $f_m = \norm{\mathbf{v}}/\lambda$;
  \item[$f_\mu$] Doppler shift for the mean \gls{doa}, i.e., $f_\mu = \hat{\mathbf{k}}_{\mu}^T \mathbf{v} / \lambda$.
\end{description}

In addition to typical mobile communication scenarios with static scatterers considered in Section~\ref{sec:ch_mod}, \eqref{eq:acf_vmf} also applies when the scattering cluster is the one being dynamic, with the \gls{tx}/\gls{rx} being static.
This is the case in radar systems, where scattering from a moving target be represented by a \gls{vmf}-distributed cluster of multipaths and \eqref{eq:acf_vmf} can be thereby used to analyze temporal correlation characteristics of the corresponding signal fluctuations at the radar  \gls{rx}.

However, for the monostatic radar case, the Doppler frequencies in \eqref{eq:acf_vmf} should be multiplied by a factor of two, to account for the fact that propagation path lengths change twice as fast compared to the change in the radar-target distance \cite{Skolnik2001book}.

\subsection{Multi-cluster Scattering}
\label{sec:corr_multiclust}
The results in \eqref{eq:scf_vmf} and \eqref{eq:acf_vmf} apply for a single cluster of \gls{vmf}-distributed scatterers.
However, their generalization to the multi-cluster scattering case typical for practical scenarios is straightforward.
As initially derived for 2D von Mises scattering in \cite{Ribeiro2005}, the \gls{scf} in channels with multiple \gls{vmf} scattering clusters can be written as%
\footnote{Here we explicitly indicate the dependence on \gls{vmf} scattering parameters, as they generally differ between clusters.} \cite{Mammasis2009}%
\begin{align}
  \hat{R}_{hh}(\mathbf{d}_{\Delta}) 
  &= \sum_{k=1}^{N_c} \hat{\Omega}_k \; \hat{R}_{hh}^k(\mathbf{d}_{\Delta}; \hat{\mathbf{k}}_{\mu, k}, \kappa_k)
  \label{eq:scf_multicl0}
\end{align}
where
\begin{description}
  \item[$N_c$] number of scattering clusters;
  \item[$\hat{\Omega}_k$] normalized power in the $k$-th cluster, i.e.,
  \begin{align}
      \sum_{k}^{N_c} \hat{\Omega}_k = 1
  \end{align}
  \item[$\hat{R}_{hh}^k$] \gls{scf} for the $k$-th cluster only, i.e., given by \eqref{eq:scf_vmf};
  \item[$\kappa_k$] spread parameter for the $k$-th cluster;
  \item[$\hat{\mathbf{k}}_{\mu, k}$] mean \gls{doa} unit vector for the $k$-th cluster.
\end{description}
The \gls{acf} is obtained similarly, by replacing the \glspl{scf} for each cluster in \eqref{eq:scf_multicl0} by the corresponding \glspl{acf}, i.e., given by \eqref{eq:acf_vmf}.

\section{Results analysis}
\label{sec:results}

\subsection{Impact of vMF Distribution Parameters on SCF}
\label{sec:results_scf_validation}
\begin{figure}[!t]
	\centering
	\subfloat[Impact of the concentration paremeter $\kappa$ ($\beta=0^\circ$).]{
		\includegraphics[scale=1]{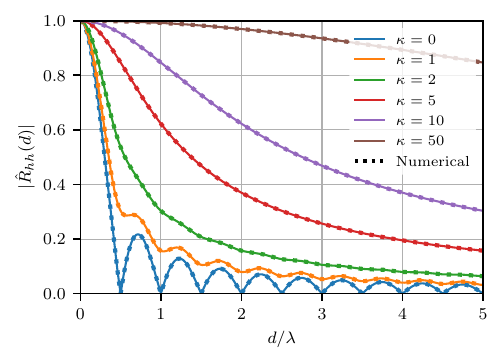}
		\label{fig:scf_vmf_th_vs_num_kappas}
	}
	\hfil
	\subfloat[Impact of displacement direction ($\kappa=10$), where $\hat{\mathbf{k}}_{\mu}^T \mathbf{d}_{\Delta} = \norm{\mathbf{d}_{\Delta}} \cos\beta$.]{
		\includegraphics[scale=1]{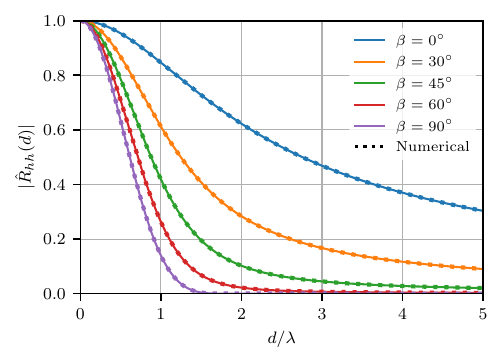}
		\label{fig:scf_vmf_th_vs_num_betas}
  }
	\caption{\gls{scf} in \gls{vmf} scattering channels, showing the impacts of (a) the concentration parameter and (b) the direction of displacement; solid lines are obtained using \eqref{eq:scf_vmf} and the dotted ones via numerical integration.}
	\label{fig:scf_vmf_th_vs_num}
\end{figure}

The impact of \gls{vmf} distribution parameters on spatial correlation is observed in Fig.~\ref{fig:scf_vmf_th_vs_num}, showing the \gls{scf} as a function of displacement distance, $d=\norm{\mathbf{d}_{\Delta}}$, for different values of the concentration parameter $\kappa$ (Fig.~\ref{fig:scf_vmf_th_vs_num_kappas}) and for different relative angles $\beta$ between the displacement direction and the mean \gls{doa} (Fig.~\ref{fig:scf_vmf_th_vs_num_betas}), where $\hat{\mathbf{k}}_{\mu}^T \mathbf{d}_{\Delta} = \norm{\mathbf{d}_{\Delta}} \cos\beta$.
Due to the parity of the cosine function, only the values of $\beta \in [0^\circ, \;90^\circ]$ need to be considered.

For $\kappa=0$, the \gls{scf} is given by \eqref{eq:scf_iso} and the characteristic oscillations of the $\sinc$ function are apparent in Fig.~\ref{fig:scf_vmf_th_vs_num_kappas}.
For higher values of $\kappa$, the \gls{scf} is complex and its magnitude falls off at a slower rate.
Therefore, spatial correlation is higher more concentrated the scatterers are.
The strong impact of the displacement direction on the \gls{scf} is observed in Fig.~\ref{fig:scf_vmf_th_vs_num_betas}.
The correlation is the highest for displacements in the direction of the mean \gls{doa}, and the lowest for the perpendicular direction.

To verify the correctness of the derived result in \eqref{eq:scf_vmf}, Fig.~\ref{fig:scf_vmf_th_vs_num} additionally shows \gls{scf} curves obtained via numerical integration (dotted lines).
The exact match is observed in all cases, thereby validating the analytically derived results.

\subsection{Spatial Correlation over Antenna Arrays}
\label{sec:results_scf_arrays}
\begin{figure}[!t]
	\centering
	\subfloat[Planar array ($\kappa=10$).]{
		\includegraphics[scale=1]{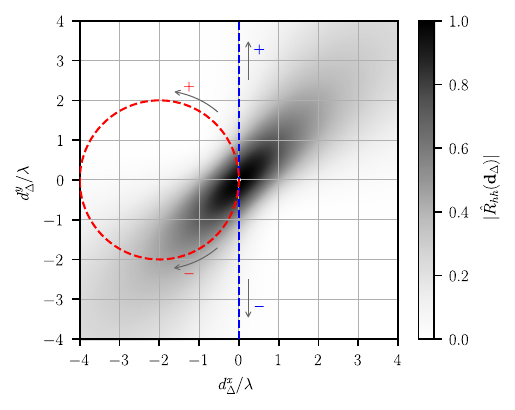}
		\label{fig:scf_vmf_array_planar}
	}
	\hfil
	\subfloat[Linear and circular arrays.]{
		\includegraphics{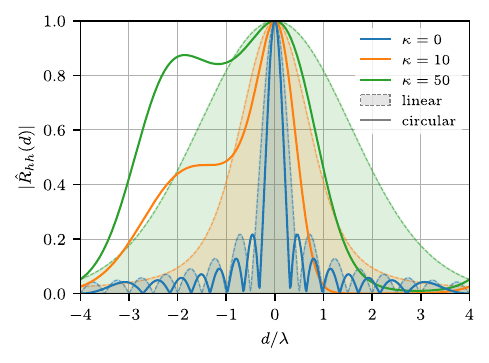}
		\label{fig:scf_vmf_array_lin_vs_circ}
  }
	\caption{\glspl{scf} for horizontal planar, linear and circular arrays; the array geometries corresponding to the results in (b) are outlined in (a).}
	\label{fig:scf_vmf_array}
\end{figure}

Fig.~\ref{fig:scf_vmf_array} shows the \gls{scf} as a function of displacement distance over planar, linear and circular arrays in the horizontal plane, for \gls{vmf} scattering with $\mu_\phi=45^\circ$ and $\mu_\psi=0^\circ$.
Fig.~\ref{fig:scf_vmf_array_planar} shows the \gls{scf} across the surface of the horizontal planar array, i.e., $\mathbf{d}_{\Delta} = (d_\Delta^x, d_\Delta^y, 0)^T$, with the dashed lines additionally indicating geometries of the considered linear (blue) and circular arrays (red).
The reference point, relative to which the correlation is calculated, is indicated by the white dot in the origin.
Fig.~\ref{fig:scf_vmf_array_lin_vs_circ} shows the \gls{scf} curves obtained for displacements along the linear and circular array geometries in Fig.~\ref{fig:scf_vmf_array_planar}.

One should note that the displacement vector along the linear array has a fixed orientation.
On the other hand, its orientation is changing for points along the circular array.
This means that the displacement distances on the horizontal axis in Fig.~\ref{fig:scf_vmf_array_lin_vs_circ} are associated with different orientations of the displacement vector for the circular array case.

As observed in Fig.~\ref{fig:scf_vmf_array_planar} for $\kappa=10$, concentrated scattering results in a higher correlation over the planar array in the direction of the mean \gls{doa} projected onto the array surface.
More concentrated scattering (higher $\kappa$) results in a wider and more elongated area of increased spatial correlation in the direction of the mean \gls{doa}.
On the other hand, for isotropic scattering ($\kappa=0$), the \gls{scf} is radially symmetric and exhibits variations according to the $\sinc$ function \eqref{eq:scf_iso} in all directions.

In consistency with the previous observations, Fig.~\ref{fig:scf_vmf_array_lin_vs_circ} shows that the signal de-correlates slower with the displacement along linear and circular arrays for more concentrated scattering.
The \gls{scf} curves obtained for the linear array are symmetric (i.e., even) and thus stationary in spatial displacement distance along the array, with the parity of the \gls{scf} being preserved regardless of the array orientation.
On the other hand, such a symmetry does not exist for the circular array and, in the considered case, a higher correlation is observed for the displacements in the negative direction, i.e., corresponding to the counter clockwise direction from the reference point along the circular path in Fig.~\ref{fig:scf_vmf_array_planar}.
Therefore, the \gls{scf} is generally non-stationary in spatial displacement along a circular array or, in general, for any non-linear one-dimensional array geometry.

\subsection{Temporal correlation of a fluctuating radar target}
\label{sec:results_acf}
\begin{figure}[t]
	\centering
	\includegraphics{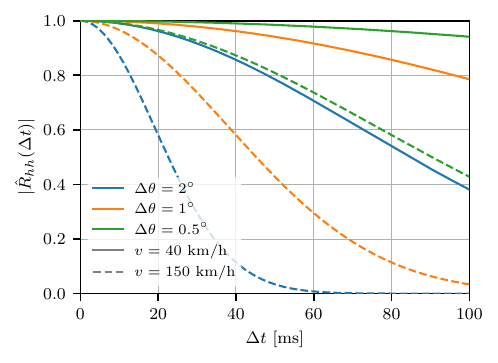}
	\caption{ACF of the \gls{rx} radar signal from moving targets with different angular widths $\Delta\theta$, observed at 20$^\circ$ elevation while traveling at a constant altitude with speeds of 40 and 120 km/h, in azimuth direction away from the radar.}
	\label{fig:results_acf}
\end{figure}

Fig.~\ref{fig:results_acf} shows the \gls{acf} of the \gls{rx} signal at a monostatic radar received for a target at an elevation angle of 20$^\circ$, represented by a \gls{vmf}-distributed cluster of scatterers, with size specified by its angular width, $\Delta \theta$.
The velocity vector of the target is assumed to be constant, horizontal and pointing in the direction away from the radar.
The operating frequency of the radar is assumed to be 10 GHz.
The choice of the concentration parameter $\kappa$ for a given $\Delta \theta$ is detailed in Appendix \ref{app:vmf_kappa}.
To circumvent the numerical instability issues described in Appendix \ref{app:scf_vmf_approx}, the approximation \eqref{eq:scf_vmf_approx} is used to evaluate the \gls{scf} for the large values of $\kappa$ in this scenario.

The results show that a faster moving target yields shorter signal de-correlation time.
By considering samples with correlation below 0.5 as being uncorrelated, for the 2$^\circ$ angular target width the signal de-correlates nearly four times faster for target speed of 150 km/h (24\,ms) than 40 km/h (85\,ms).
The target size also plays an important role, where the \gls{rx} signal de-correlates slower for smaller targets and faster for the larger ones.
For the considered target speed of 150\,km/h, the angular target widths of 2$^\circ$, 1$^\circ$ and 0.5$^\circ$ yield de-correlation times of 24\,ms, 46\,ms and 90\,ms, respectively.

It is important to point out that the direction of target motion also has a significant impact on the \gls{rx} signal de-correlation which is not considered here, but can be anticipated based on the results in Fig.~\ref{fig:scf_vmf_th_vs_num_betas}.
The \gls{rx} signal de-correlation with the change in the target's orientation is neglected here, but has an important impact in practical radar systems.
Similarly, the frequency correlation needs to be appropriately modelled in systems employing frequency agility techniques \cite{Skolnik2001book}.

\section{Conclusions}
\label{sec:conclusions}
The \gls{vmf} distribution is a widely adopted model for representing clusters of scatterers in wireless communication channels, taking both the main direction and the concentration of scattering into account.
This letter presents a simple closed-form expressions for the \gls{scf} and \gls{acf} in channels with \gls{vmf} scattering.
The simplicity of the presented expressions provides a better insight into the impact of the \gls{vmf} distribution parameters on spatial correlation properties, and allows for analysis of arbitrary antenna array structures.

The correctness of the obtained expressions is verified by comparison against numerical integration results, where an exact match is observed.
Together with the repeated experimental demonstration of the validity of the \gls{vmf} scattering model in the literature, this agreement establishes practical relevance of the presented results.

To demonstrate their utility, the derived expressions are employed to analyze spatial correlation across planar, linear and circular arrays.
The results show that concentrated scattering yields a higher spatial correlation in the direction of the mean \gls{doa}, increasing with higher values of $\kappa$.
The non-stationarity of the \gls{scf} in displacement distance for circular and other non-linear array geometries is observed.
The obtained results are also used to investigate temporal correlation characteristics of a radar signal reflected from a moving target, represented by a cluster of \gls{vmf}-distributed scatterers, with the results showing the important impact of the target size, speed and motion direction on the \gls{rx} signal fluctuation dynamics.

\appendices

\section{}
\label{app:vmf_kappa}
The value of $\kappa$ corresponding to a given angular width of the target in Sec.~\ref{sec:results_acf} is obtained according to
\begin{align}
  \kappa(\Delta\theta) = \frac{2}{1 - \cos\left(\frac{\Delta\theta}{2}\right)}
\end{align}
where
\begin{description}
  \item[$\Delta\theta$] angular width of the target seen by the radar.
\end{description}
This expression is obtained by considering the angular deviation from the mean \gls{doa}, for which the \gls{vmf} \gls{pdf} falls to a fraction $e^{-2} \approx 0.13$ of its maximum value.

\section{}
\label{app:scf_vmf_approx}
Due to the numerical instability of the available algorithms for evaluation of the $\sinh$ function for very large values of the argument, one encounters issues when using \eqref{eq:scf_vmf} to evaluate the \gls{scf} for very large $\kappa$ (i.e., typically for $\kappa>700$).
In this case, the following tight approximation is useful
\begin{align}
  \hat{R}_{hh}(\mathbf{d}_{\Delta})
  &\approx \kappa e^{-\kappa} \frac{e^{jz}}{jz}
  \label{eq:scf_vmf_approx}
\end{align}
where $z$ is the argument of the \gls{scf} in \eqref{eq:scf_vmf}, i.e.
\begin{align}
  z = \sqrt{\left(\frac{2\pi}{\lambda}\right)^2 \norm{\mathbf{d}_{\Delta}}^2 - \kappa^2 - j \frac{4\pi\kappa}{\lambda} \, \hat{\mathbf{k}}_{\mu}^T \mathbf{d}_{\Delta} }
\end{align}
This approximation also applies to \eqref{eq:acf_vmf}, with appropriately replaced argument $z$.

\ifCLASSOPTIONcaptionsoff
  \newpage
\fi



\end{document}

%% file: acronyms.tex
\newacronym[\glslongpluralkey={Directions of Departures}]{dod}{DoD}{Direction of Departure}
\newacronym[\glslongpluralkey={Directions of Arrivals}]{doa}{DoA}{Direction of Arrival}
\newacronym[\glslongpluralkey={Angles of Departure}]{aod}{AoD}{Angle of Departure}
\newacronym[\glslongpluralkey={Angles of Arrival}]{aoa}{AoA}{Angle of Arrival}
\newacronym[\glslongpluralkey={Elevation Angles of Departures}]{eaod}{EAoD}{Elevation Angle of Departure}
\newacronym[\glslongpluralkey={Elevation Angles of Arrivals}]{eaoa}{EAoA}{Elevation Angle of Arrival}
\newacronym[\glslongpluralkey={Azimuth Angles of Departures}]{aaod}{AAoD}{Azimuth Angle of Departure}
\newacronym[\glslongpluralkey={Azimuth Angles of Arrivals}]{aaoa}{AAoA}{Azimuth Angle of Arrival}
\newacronym{mpc}{MPC}{Multipath Component}
\newacronym[first=transmitter (Tx)]{tx}{Tx}{Transmitter}
\newacronym[first=receiver (Rx)]{rx}{Rx}{Receiver}
\newacronym{vmf}{vMF}{von Mises-Fisher}
\newacronym{fb5}{FB5}{Fisher-Bingham}
\newacronym{pdf}{PDF}{Probability Density Function}
\newacronym{cdf}{CDF}{Cumulative Distribution Function}
\newacronym{scf}{SCF}{Spatial Correlation Function}
\newacronym{acf}{ACF}{Auto-Correlation Function}
\newacronym{ccf}{CCF}{Cross-Correlation Function}
\newacronym{psd}{PSD}{Power Spectrum Density}
\newacronym{mimo}{MIMO}{Multiple-Input Multiple-Output}
\newacronym{icas}{ICAS}{Integrated Communication and Sensing}
\newacronym{iid}{i.i.d.}{independent identically distributed}
\newacronym{uav}{UAV}{Unmanned Aerial Vehicle}